# ARTICLE

# Density fluctuations and border forces direct leader cell plasticity during collective epithelial migrations

Sameeksha Rao[a], Suvakash Dey[a] and Namrata Gundiah*[a]

Epithelial cell monolayers expand on substrates by forming finger-like protrusions, created by leader cells, in the monolayer boundary. Information transmission and communication between individual entities in the cohesive collective lead to long-range order, vortical structures, and disorder to ordered phase transitions. We ask the following questions: what makes a leader? What is the role of followers in leader cell formation? We used a particle-based model to simulate epithelial cell migrations on substrates of 9.4 kPa, 21 kPa and 33 kPa stiffness. The dynamics of cellular motion in the ensemble are governed by orientational Vicsek and inter-cellular interactions between neighboring particles. The model also includes bending, curvature-based motility, and acto-myosin contractile cable forces on the contour, in addition to density dependent noise and cell proliferations. We show that border forces are essential in the leader cell formation and the overall areal expansions of epithelial monolayers on substrates. Radial velocities and areal expansions of the monolayer agree with experiments reported for epithelial cells on substrates of varied stiffness. Ordering in follower cells within a specific region of the monolayer was apparent on substrates of higher stiffness and occurred prior to the emergence of leader cells. We demonstrate that regions of increased cell density occur behind the leader cell edge on all three substrates. Finally, we assessed the role of cell divisions on the ordering of velocities in the monolayer. These results demonstrate that monolayer heterogeneities, caused by density instabilities in the interior regions, correlate with leader cell formation during epithelial migrations.

## Introduction

Collective epithelial migrations occur in streams, sheets, and clusters during development, normal wound healing, and invasive movements of cancer cells[1,2]. Confluent cells in epithelial monolayers attach to substrates through transmembrane integrin proteins, located in focal adhesions, and crawl using coordinated and cooperative active traction generating mechanisms to display solid-like behaviors in the short time interval and fluid-like viscous behaviors over longer durations[3,4]. The key mechanisms driving collective epithelial migrations include lamellipodial extensions in cells located at the leading cell boundary, actin polymerization, actomyosin contractility, cell substrate adhesions, and cell-cell connections[5]. Migrating cells in epithelial layers create finger-like projections at the leading edge of the ensemble, called leader cells, and are accompanied by follower cells in the monolayer interior [6,7].

Leader cells have clear front-rear polarizations, exhibit lower contact inhibition, are more propulsive as compared to interior cells, and direct migrations[8]. Rac-dependent actin polymerization promotes lamellipodial and filopodial extensions in the front end whereas antagonistic Rho activated myosin II dependent mechanisms induce contractile forces at the rear end of polarized cells[6,8]. The physical basis for the origin of leader cells in epithelial ensembles has been of some debate[9]. Dynamic instabilities, associated with monolayer elasticity and bounding edge curvatures, aid in the creation of leader cells[7,10]. The presence of a tensile actomyosin cable in the boundary acts as a purse-string to exert large forces on the monolayer boundary; disruptions in cable forces are important in leader cell creation[10,11,12]. Leader cells in endothelial monolayers express different genes as compared to interior cells in the collective[13]. External cues, neighbor interactions, and other soluble factors are also hypothesized to be important in the formation of leader cells[14].

Force transmission in the monolayer occurs through transmembrane mediated adhesions that link the actomyosin cortexes of neighboring cells to permit shearing between cells during migration[4]. Frictional protein-mediated forces on the substrate balance active tractions generated by cells[15]. Traction maps show a highly heterogeneous stress field that operates over many cell lengths starting from the leading cell edge of the monolayer[16]. Increased tractions occur behind prospective leaders; followers hence exert forces and aid in the creation of leader cells[17]. Sepulveda and co-workers used a computational model to simulate instabilities in epithelial sheets by imparting different properties to leader cells[18]. Other studies suggest that local curvatures in the monolayer are critical in leader cell

[a.] Biomechanics Laboratory, Department of Mechanical Engineering, Indian Institute of Science, Bangalore 560 012.
*Email: namrata@iisc.ac.in; ngundiah@gmail.com

formation through changes in the actomyosin cable force[7,12]. Continuum polar fluid models of epithelial monolayers show that active cellular tractions primarily contribute to the fingering instability through sustained generation of a velocity gradient[9]. Finger generation through this kinematic mechanism does not require assigning specific properties to leader cells or regulation of the local monolayer curvatures.

Computational particle-based models are useful to quantify the intricate mechanisms that lead to clustering, self-assembly, and pattern formation in dynamical systems of self-propelled particles such as epithelial cell ensembles in a monolayer[5]. The Vicsek model, and the interaction forces between particles, are both essential features of such mathematical models[19,20]. Particle-based models for collective epithelial migrations demonstrate streaming and swirling coordinated motion[12,18]. Border and actomyosin cable forces are essential in quantifying the moving fronts in epithelial migrations based on experiments and models[7,10,12,18,21]. Balcioglu and co-workers patterned Madin–Darby canine kidney (MDCK) epithelial cells on polyacrylamide gels of various stiffness and showed increasing radial velocities on higher stiffness substrates[22]. In contrast, the number of finger-like projections associated with leader cells was highest on intermediate substrate stiffness.

We ask the following questions: what makes a leader? What is the role of follower cells in the formation of leader cells? We use a particle-based model of epithelial cells on substrates of varied stiffness to investigate differences in migration behaviors and assess the physical factors that contribute to the leader cell formation on substrates of different stiffness. The model includes velocity alignment and force interactions between neighbor cells, in addition to contour forces and cell proliferations on the substrates. We show the importance of border forces in leader cell creation. We demonstrate increased particle ordering on higher stiffness substrates and ordering in the velocities of follower cells during monolayer expansion. We also show regions of increased cell density that occur behind the leader edge on all substrate stiffnesses prior to leader cell formation. Finally, we assess the role of cell divisions in the monolayer that may contribute to the ordering of velocities and leader cell creation during epithelial cell migrations.

## Model Development and Calibration

We use a particle ensemble formulation, based on earlier literature, and define each cell in the monolayer using their centroids[12,18,19,20,21,23]. The dynamics of cellular motion are governed by orientational Vicsek and inter-cellular interactions between neighboring particles in the ensemble[19,20]. Vicsek interactions in an assembly of particles, moving with a constant velocity in discrete time steps, result in particle alignment along a specific direction within a neighborhood of the particle under consideration[19]. Ensemble models also include a Lennard-Jones type of body force that operates over a short cohesion length between each pair of particles[20]. These inter-cellular forces relate to the attraction between individual particles, repulsion at very close distances, and cooperativity through interplay to ensure collective behaviors. A density dependent noise term, included in the model for random perturbations, accounts for ordering during collective migrations. The dynamical equation for each cell, $i$, in the ensemble is given based on the velocity, $v_i$, as

$$\frac{dv_i}{dt} = -\alpha v_i + \sum \left[\frac{\beta}{N_i}(v_j - v_i)\right] + \sum[F_{int}(r_{ij})] + \sigma(\rho)\eta_i \quad (1)$$

The term $-\alpha v_i$ describes friction between the cell and the underlying substrate. Summation in the equation is on $N_i$ neighbors of each cell, i, with neighbor cell, j. The parameter, $\beta$, dictates the orientational ordering between velocities of the cell, $i$, with the neighboring cell, $j$, located at a separation distance, $r_{ij}$. The force $F_{int}$ describes the attractive-repulsive interaction between a pair of cells, i and j, and is given by[12]

$$F_{int} = U_o r \exp\left(-\left(\frac{r_{ij}}{A_o}\right)^2\right) + U_1(r_{ij} - A_1)H(r_{ij} - A_1) + U_2 \exp\left(-\frac{r_{ij}}{A_2}\right) - U_3(r_{ij} - A_3)^2 H(r_{ij} - A_3) \quad (2)$$

Model parameters, $U_k$, and $A_k$; $k = 0:3$, in the equation define the strength of attractive or repulsive forces that depends on the distance of separation ($r_{ij}$) between a pair of cells. The last term in eqn (1) is the random noise, included as an Ornstein–Uhlenbeck process, in the system. The magnitude of noise is dependent on the local density ($\rho$) and is given by

$$\sigma(\rho) = \sigma_o + (\sigma_1 - \sigma_o)\left(1 - \frac{\rho}{\rho_o}\right) \quad (3)$$

The amplitude of noise increases with a decrease in local density. The direction of application of noise ($\eta_i$) for a given cell (i) changes with time and is given in terms of $\tau$, the correlation time, and $\xi_i$, the delta-correlated white noise, as,

$$\tau \frac{d\eta_i}{dt} = -\eta_i + \xi_i \quad (4)$$

Three additional contour forces act on cells in the monolayer border. First, a restorative bending force, $F_{bend}$, which depends on the curvature ($H$), due to elasticity of the monolayer at the border, and is given by

$$F_{bend} = -\kappa \frac{d^2 H}{ds^2} + \frac{3}{2}\kappa |H^2| H \quad (5)$$

Second, an outward directed motility force that depends on the vector value of convex curvature, $H$, of the membrane and is given by,

$$F_{border} = \begin{cases} F_o & H > 0 \\ \frac{F_{max}}{H_{max}}|H| & 0 > H > -H_{max} \\ F_{max} & H < -H_{max} \end{cases} \quad (6)$$

Finally, cable forces ($F_{cable}$) of a constant magnitude act on the concave portions of the boundary. $F_{cable}$ represents the tensile force due to the presence of an acto-myosin cable that is observed in cells located in concave and flat regions of the monolayers[10,12]. We define a leader cell in this study as one with a convex curvature lesser than $-H_{max}$ (Table 2).

Cell proliferation is implemented in the model as a term which depends on the local density and the division time for MDCK cell cultures, $T_o$, given as [21]

$$T(\rho) = T_o \left(1 + \left(\frac{\rho}{\rho_o}\right)^4\right) \qquad (7)$$

$\rho_o$ is the reference density at the time of cell seeding and $\rho$ the local density at each of the different times during migration.

We used parameters for MDCK cells in the particle-based model that best matched the experimental results presented by Balcioglu and colleagues on substrates of 9.4 kPa, 21 kPa and 33 kPa substrate stiffness[22]. Cells were initially seeded in a circular pattern on each of the substrates with areas reported in the study (Fig. 1; Table 1). Fig. 1a shows cells on the 21 kPa substrate stiffness at 16 hours during migration. Cells proliferate and contribute to an increase in the monolayer area. Initially seeded cells in the model are shown in blue, new cells obtained due to division are represented in green, and leader cells in the monolayer boundary are indicated using magenta. To identify leader cells, we calculated the tangent vector at each cell on the border to obtain the curvature. A cell was defined as a leader cell if the resultant curvature was negative (convex), and the magnitude exceeded $H_{max}$ (Table 2). Distances and angles between neighboring cells were computed and are shown for a region indicated in the subplot of Fig. 1a. In this method, the neighborhood of each cell was divided into equal sectors of 60º and the nearest cell in each sector from a given cell was identified as the nearest neighbor (shown in red)[18]. Probability distributions for the distances between individual particles, and the angles between neighbors, were fit to normal distributions and the values compared with experimentally quoted results at the start of the experiment (Fig. 1b and Fig. 1c)[22]. Model parameters, $F_{cable}$ and $F_{max}$, were varied to match experimentally reported monolayer areas and the percentage of leader cells at 16 hours. Parameters for inter-cellular interaction, ($U_i$ and $A_i$), were obtained such that the leader cells did not detach from the monolayer. The terms $\beta$, and noise parameters, $\sigma_o$ and $\sigma_1$, in the model were selected to obtain experimentally reported radial velocity and PIV velocity between 50-100 $\mu m$ behind the border for each substrate. A total of 6 different simulations were performed for each value of the different parametric variations; mean and standard deviations from these simulations are included in all comparisons. Table 2 shows the final parameters used in the model for the three different substrate stiffnesses.

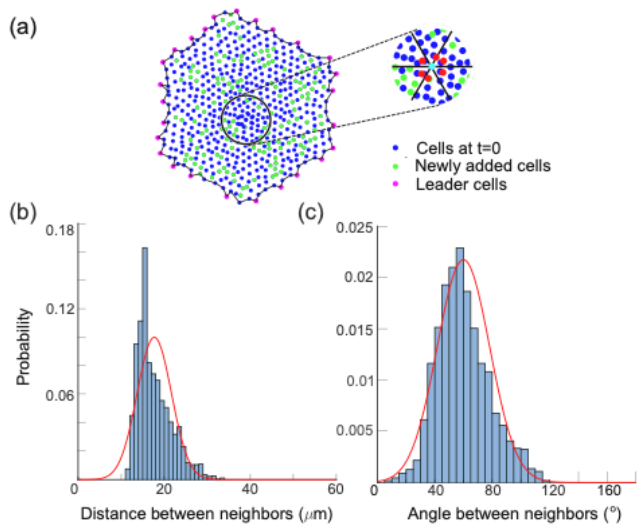

**Figure 1:** (a) The leader cells (magenta), new cells due to proliferation (green), and the nearest neighbors (red) of a give cell (cyan) are shown at 16 hours for the 21 kPa substrate (b) The probability distribution for the distance between each cell and its neighbors in the monolayer is shown at time of seeding (0 hours). (c) The probability distribution for the angles between each cell and its neighbors in the monolayer is shown at 0 hours.

Table 1: Cell Seeding Parameters on the Substrates

| Substrate Stiffness (kPa) | Monolayer Diameter ($\mu m$) | Number of Cells at T = 0 hr | Distance between Neighboring Cells ($\mu m$) | Angle between Neighboring Cells (º) |
|---|---|---|---|---|
| 9.4 | 331 | 325 | 19.56 ± 4.84 | 60 ± 19.48 |
| 21 | 356 | 444 | 17.69 ± 3.99 | 60.05 ± 18.65 |
| 33 | 354 | 406 | 18.50 ± 4.26 | 60.03 ± 18.93 |

Table 2: Model Parameters

| Parameter | Notation | Value | Reference |
|---|---|---|---|
| Short-Range Repulsion | $U_o$ (hr$^{-2}$) | 2650 | (ref. 12) |
| | | 660 | (9.4, 21 and 33 kPa) |
| | $A_o$ ($\mu m$) | 8 | (ref. 12) |
| | | 7 | (9.4, 21 and 33 kPa) |
| | $U_2$ ($\mu m$ hr$^{-2}$) | 2 | (ref. 12, 9.4, 21 and 33 kPa) |
| | $A_2$ ($\mu m$) | 25 | (ref. 12, 9.4, 21 and 33 kPa) |
| Long-Range Attraction | $U_1$ (hr$^{-2}$) | 30 | (ref. 12) |
| | | 3 | (9.4, 21 and 33 kPa) |
| | $A_1$ ($\mu m$) | 2 | (ref. 12, 9.4, 21 and 33 kPa) |
| | $U_3$ ($\mu m^{-1}$ hr$^{-2}$) | 1 | (ref. 12) |
| | | 0.6 | (9.4 and 33 kPa) |
| | | 2 | (21 kPa) |
| | $A_3$ ($\mu m$) | 26 | (ref. 12) |
| | | 10 | (9.4, 21 and 33 kPa) |
| Friction Coefficient | $\alpha$ (hr$^{-1}$) | 1.42 | (ref. 12) |
| | | 21.3 | (9.4, 21 and 33 kPa) |
| Correlation Time | $\tau$ (hr) | 1.39 | (ref. 12, 9.4, 21 and 33 kPa) |
| Orientational Interactions | $\beta$ (hr$^{-1}$) | 60 | (ref. 12) |
| | | 30 | (9.4 and 21 kPa) |
| | | 50 | (33 kPa) |
| Proliferation Parameter | $T_o$(hr) | 35 | (ref. 12) |
| | | 28 | (9.4 and 33 kPa) |
| | | 22 | (21 kPa) |
| Bending Modulus | $\kappa$ ($\mu m^4$ hr$^{-2}$) | 2.5 | (ref. 12, 9.4, 21 and 33 kPa) |
| Cable Forces | $F_{cable}$ ($\mu m$ hr$^{-2}$) | 350 | (ref. 12) |
| | | 200 | (9.4 kPa) |
| | | 800 | (21 and 33 kPa) |
| | | 1250 | (ref. 12) |

| Maximal Motile Force | $F_{max}$ ($\mu m\ hr^{-4}$) | 1600 | (9.4 kPa) |
|---|---|---|---|
| | | 1900 | (21 and 33 kPa) |
| Maximal Curvature | $H_{max}$ ($\mu m^{-1}$) | 0.05 | (ref. 12) |
| | | 0.045 | (9.4, 21 and 33 kPa) |
| Noise | $\sigma_o$ ($\mu m\ hr^{-\frac{3}{2}}$) | 150 | (ref. 12) |
| | | 200 | (9.4 kPa) |
| | | 300 | (21 kPa) |
| | | 250 | (33 kPa) |
| | $\sigma_1$ ($\mu m\ hr^{-\frac{3}{2}}$) | 300 | (ref. 12) |
| | | 240 | (9.4 kPa) |
| | | 360 | (21 kPa) |
| | | 400 | (33 kPa) |
| Reference Density | $\rho_o$ ($\mu m^{-2}$) | 0.0022 | (ref. 12) |
| | | 0.004 | (ref. 18, 9.4, 21 and 33 kPa) |
| Gaussian Width | $\sigma_G$ ($\mu m$) | 35 | (ref. 12, 9.4, 21 and 33 kPa) |

## Results and Discussion

Synchronized movements in birds, fish, bacterial swarms, self-propelled matter, and epithelial cells arise from individual and coordinated interactions in the collective that result in emergent behaviors. Information transmission and communication between individual entities in the cohesive collective lead to long-range order, vortical structures, and disorder to ordered phase transitions[7,10,12,24]. There are many similarities in these diverse and complex systems, despite obvious size scales differences, that suggest the existence of universal organizing principles in collective migrations[25]. Directed migrations in epithelial monolayers occur through the formation of leader cells[25]. We used a particle-based model to simulate leader cell formation on substrates of 9.4 kPa, 21 kPa and 33 kPa stiffness. There are three main findings from this study: first, we show that border forces are important in the creation of leader cells. Second, ordering in follower cells within the monolayer is apparent on substrates of higher stiffness and occurs before the emergence of a leader cell on all three substrate stiffnesses. Finally, heterogeneous regions of increased cell density occur behind the leader cell edge which demonstrates the importance of density instabilities in leader cell formation in the epithelial monolayer.

**Border forces play a key role in leader cell formation**

Border forces at lamellipodia, and due to the presence of an acto-myosin contractile cable, are hypothesized to be essential drivers of collective migration in epithelial monolayers[10,12,26]. To quantify the role of border forces in leader cell formation during migration, we parametrically varied values of the border forces in the model and compared results with experiments for the 33 kPa substrate stiffness. Balcioglu and co-workers reported intermediate values in the percentage of leader cells on the 33 kPa substrate relative to the 9.4 kPa and the 21 kPa substrate; in contrast, radial velocities were highest on the 33 kPa substrate stiffness[22]. Fig. 2 shows the percentage of leader cells at 16 hours of migration on 33 kPa substrate. Fig. 2a was obtained by parametrically varying the border force, $F_{max}$, over a range of values (1600 – 2500 $\mu m\ hr^{-4}$) for a constant cable force ($F_{cable}$=800 $\mu m\ hr^{-2}$). We note a decrease in the percentage of leader cells at high values of $F_{max}$ in simulations due to cell detachment from the monolayer. For cases where detachments do not occur, the percentage of leaders increased with $F_{max}$. ANOVA tests with Bonferroni comparisons show that these values were however not statistically significant (p<0.05). Fig. 2b illustrates the effects of increasing cable force ($F_{cable}$) on the percentage of leader cells for a constant value of border force ($F_{max}$=1900 $\mu m\ hr^{-4}$). There are no significant differences in the percentage of leader cells formed with increase in cable force over a range of values from 100 $\mu m\ hr^{-2}$ to 900 $\mu m\ hr^{-2}$ (p<0.05). These studies show that the border force, $F_{max}$, is critical in determining the percentage of leader cells formed during migration. Border forces occur due to protrusive forces at the lamellipodia that drive collective migrations[26]. Such forces are not only important in leader cell formation, but also contribute to the monolayer expansion. Higher values of border forces result in higher rates of monolayer expansion[12].

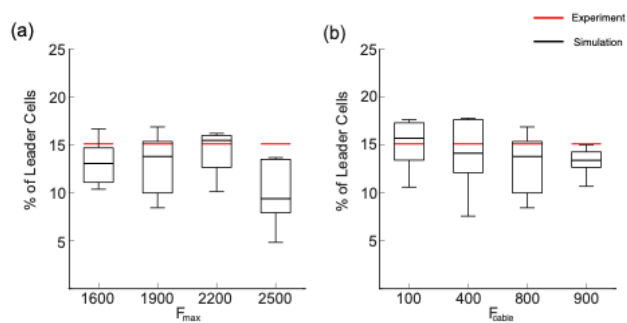

Figure 2: (a) Percentage of leader cells formed during migration are plotted at 16 hours for varying border forces ($F_{max}$) at constant cable force ($F_{cable}$= 800) (b) The percentage of leader cells are plotted at 16 hours for varying cable force ($F_{cable}$) at constant cable force ($F_{max}$= 1900). The box plots represent an average of 6 simulations. Experimentally reported values are indicated in the figure in red.

We next explored particle migrations on the 9.4 kPa and 21 kPa substrates using optimized values of $F_{max}$=1900 $\mu m\ hr^{-4}$ and $F_{cable}$=800 $\mu m\ hr^{-2}$ and compared these results with experimentally reported values for percentage of leader cells and monolayer area at 16 hours. We note a higher percentage of leader cells on the 21 kPa substrate stiffness for these values of border and cable forces. Because leader cells and the monolayer areas were both significantly lower on the 9.4 kPa substrate as compared to the 33 kPa substrate, we changed the border forces $F_{max}$ and $F_{cable}$ to 1600 $\mu m\ hr^{-4}$ and 200 $\mu m\ hr^{-2}$ respectively to match experimental data[22]. We also varied the inter-cellular interaction terms in eqn (2) for the 21 kPa substrate such that a higher percentage of leader cells could be formed and sustained without cells detaching from the monolayer. This allowed an increase in the long-range attraction forces, governed by terms $U_1$, $A_1$, $U_3$ and $A_3$ in eqn (2), and a decrease in the short-range repulsion forces that are determined by $U_o$, $A_o$, $U_2$ and $A_2$ between neighboring cells. Because of a change in the value of the interaction potential, the monolayer expansion on the 21 kPa substrate also decreased for these values of $F_{max}$ and $F_{cable}$.

Fig. 3a compares results of the simulation with experiments for the percentage of leader cells formed at 16 hours, and Fig. 3b demonstrates the normalized area of the monolayer at the same time point for all three substrates[22]. The highest percentage of leader cells were formed on the substrate with intermediate stiffness (21 kPa) with no significant difference between the percentage of leader cells formed on the 9.4 kPa and 33 kPa substrates respectively. The 33 kPa substrate had the greatest normalized monolayer area at 16 hours; the normalized areas for 9.4 kPa and 21 kPa substrates were however similar.

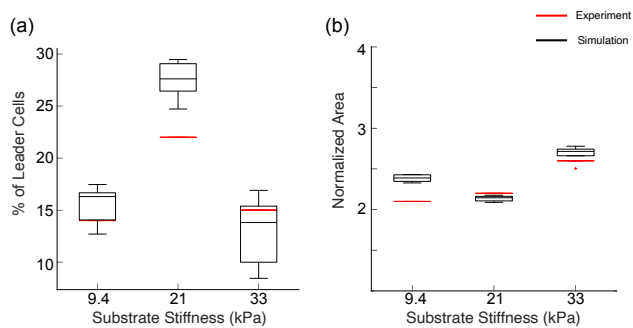

**Figure 3:** Results from experiments (red) and the particle-based model (black) are shown for (a) the percentage of leader cells, and (b) the monolayer area, normalized to initial area, at 16 hours for all three substrates. The box plots represent an average of 6 simulations. There is good agreement between the simulations and experimentally reported values for the different substrate stiffness values using the tuned model parameters.

We compared the radial and PIV velocities from simulations with experimentally reported values that occur 50-100 $\mu m$ behind the border at 16 hours. Tarle and co-workers showed that the alignment parameter ($\beta$) and density dependent noise ($\sigma_1$ - $\sigma_o$) are both essential to obtain a suitable velocity profile in follower cells during migration[12]. Because the PIV and radial velocities of cells were highest on the 33 kPa substrate, higher values of the alignment parameter ($\beta$=50 h$r^{-1}$), and parameters contributing to the density dependent noise terms, $\sigma_o$ =250 $\mu m$ hr$^{-\frac{3}{2}}$ and $\sigma_1$ =400 $\mu m$ hr$^{-\frac{3}{2}}$, were required to match cell velocities in the region behind the border. PIV and radial cell velocities on the 21 kPa substrate matched experimental results for lower values of the alignment parameter ($\beta$=30 h$r^{-1}$) and density dependent noise ($\sigma_o$ =300 $\mu m$ hr$^{-\frac{3}{2}}$ and $\sigma_1$ =360 $\mu m$ hr$^{-\frac{3}{2}}$) relative to the 33 kPa substrate. A combination of $\beta$ =30 h$r^{-1}$, $\sigma_o$ =200 $\mu m$ hr$^{-\frac{3}{2}}$ and $\sigma_1$ =240 $\mu m$ hr$^{-\frac{3}{2}}$ matched the experimental values for the 9.4 kPa substrate that had the lowest velocities (Fig. 4a; Table 2). Supplementary video 1 illustrates the evolution of the cell velocities and vorticities on the 21 kPa substrate throughout the 20-hour migration period.

**Formation of leader cells in epithelial cell monolayers**

We identified and tracked individual leaders during migration on substrates with variable stiffness. Fig. 5 shows cell trajectories, following identification of leader cells, over the total migration duration (20 hours) for all substrates. The trajectories of border cells are marked black before cells become leaders, and in green after the cells become leaders. Interior cells trajectories are tagged blue before they become leaders, and red after they become leaders. These results show that the migration directions of leader cells are relatively straight on 9.4 kPa and 33 kPa substrate. In contrast, cell trajectories on the 21 kPa substrate were more tortuous due to the increased noise and a lower alignment parameter in the system (Table 2). Most leader cells in the monolayer are in the monolayer border as compared to cells that were in the interior regions.

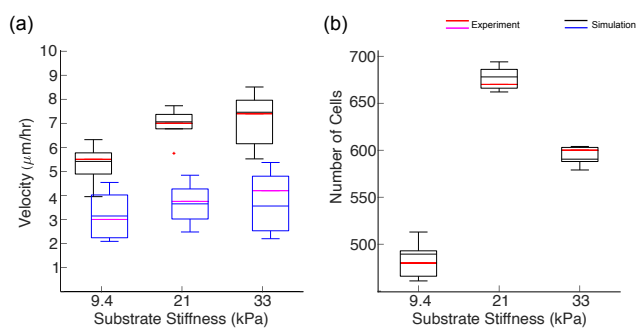

**Figure 4:** (a) Results from experiments (red) and the particle-based model (black) were compared for the radial and PIV velocities from cells located 50-100 μm behind the border at 16 hours for all three substrates. (b) Comparisons in the number of cells reported in experiments (red) and those obtained from the particle-based model (black) are shown at 16 hours for all three substrates. The box plots represent an average of 6 simulations.

Vishwakarma and co-workers suggested the importance of increased dynamic heterogeneity, related to cell densities in the monolayer bulk, that facilitate efficient collective migration[27]. Little is however understood if the dynamics and densities of follower cells play a direct role in leader cell creation. We quantified the time of formation of leader cells in the model by tracking individual leader cells in the collective and explored the contributions of follower dynamics in this process. The order parameter, defined as the degree to which the cell velocity aligns in the radial direction, was quantified for monolayer migrations on substrates of varied stiffness using the particle based model[28]. Fig. 6 shows plots of the order parameter in the monolayer at 15, 30, 60, and 120 minutes before a representative cell becomes a leader cell. A circular region of radius 100 $\mu m$ was defined to quantify the ordering of followers on all three substrates. The monolayer interior has lower ordering as compared to regions at the edges for all substrate stiffness. Followers are also ordered within the periphery of the monolayer up to 120 minutes before the emergence of a leader cell on all substrates. This ordering primarily occurs due to the border force in the monolayer boundary which is a function of the magnitude of convex curvature at the border. A lower value of convex curvature is present prior to a border cell becoming a leader cell. A smaller

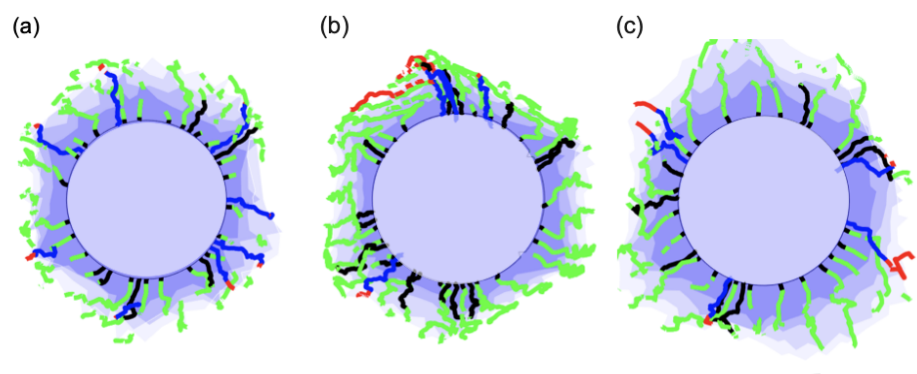

**Figure 5:** The trajectories of leader cells during the 20-hour migration period are plotted for the (a) 9.4 kPa, (b) 21 kPa and (c) 33 kPa substrates. The green-black trajectories are paths of leader cells that were border cells prior to their emergence as leaders. The red-blue trajectories are paths of leader cells that were interior cells prior to their emergence as leaders. Scale bar: 170 $\mu$m

magnitude of border force is hence present at that cell based on eqn (6). This smaller magnitude of border force imparts directionality to the border cell, through Vicsek interactions, and induces order among the follower cells. Fig. 6 also shows the level of order that depends on substrate stiffness. The highest degree of ordering was observed for the stiffest substrate (33 kPa) whereas the lowest ordering was visible for the 9.4 kPa substrate. The areal expansion of the monolayer also showed a similar dependence on substrate stiffness. A higher ordering and directionality among the cells in the interior correlated with a greater rate of monolayer expansion. Supplementary video 2 illustrates the variations in the order parameter on the 33 kPa substrate throughout the 20-hour migration period.

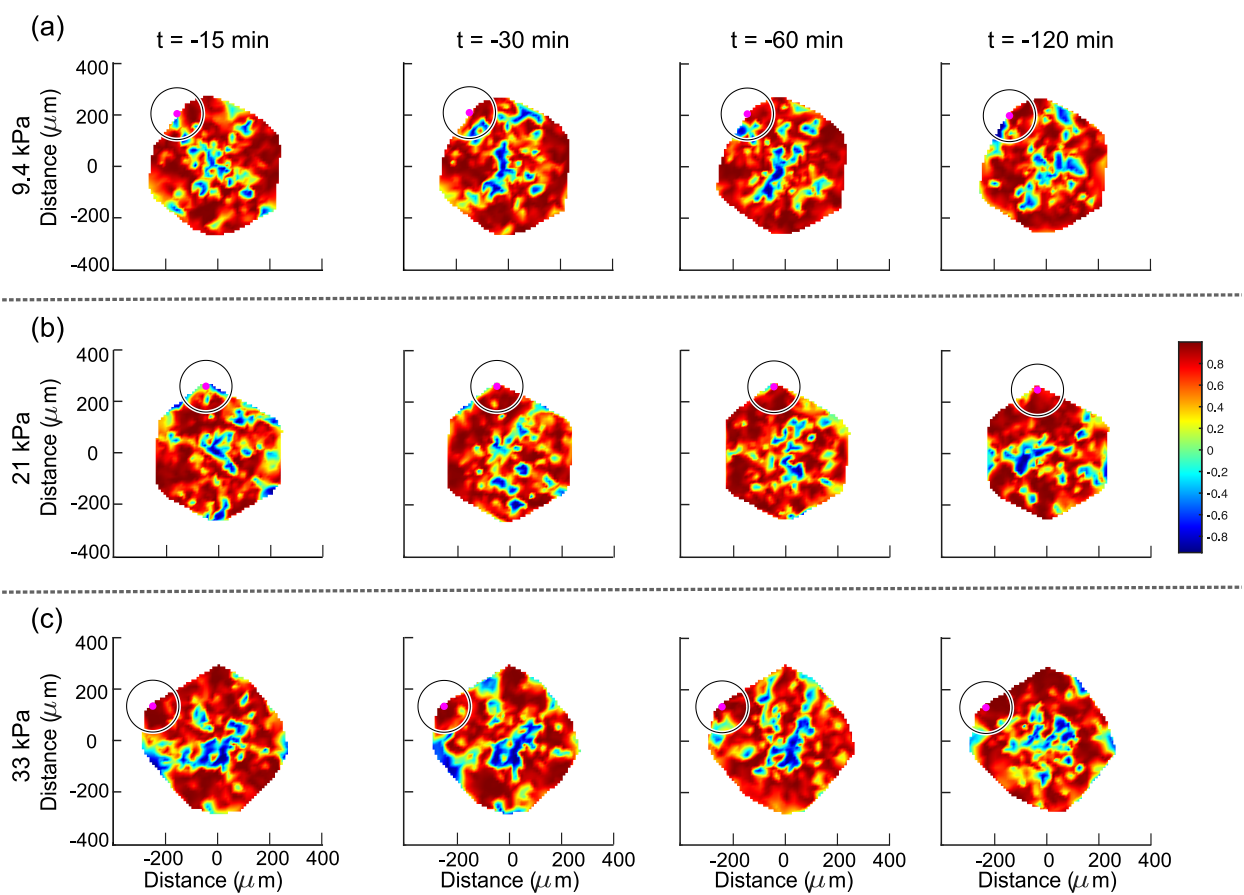

**Figure 6:** Order parameter plots are shown for the (a) 9.4 kPa substrate, (b) 21 kPa substrate and (c) 33 kPa substrate at 15 mins, 30 mins, 60 mins and 120 mins prior to a cell (magenta) becoming a leader. The 33 kPa has the highest ordering, 21 kPa intermediate ordering and 9.4 kPa has the least ordering. The circle defines a 100 μm region around the concerned cell (magenta).

**Contributions from cell proliferations and increased density to leader cell formation**

Mechanical coupling between cells, and cell-substrate interactions influence cell proliferations on substrates. Cells have increased cell areas during their proliferation stage and during expansion from an initial confinement[29]. The number of cells in our model was determined using a proliferation parameter, $T_o$, in eqn (7). Balcioglu and co-workers showed a maximum number of cells on the 21 kPa substrate relative to the 9.4 kPa and 33 kPa substrate[22]. The average division time is dependent on the local density and given using eqn (7) [30]. We used $T_o$ = 22 hr for the 21 kPa substrate and $T_o$ = 28 hr for the 33 kPa and 9.4 kPa substrates respectively in this study to match simulations with experimentally reported results. Fig. 4b shows good agreement between the results from the particle-based model and the experimental results for the number of cells in the monolayer at 16 hours for all three substrates.

We next calculated the divergence of cell velocities at various time points before and after cell division for all three substrate stiffnesses. Divergence of the velocity field shows contributions from contractions and expansions in a specific region of the monolayer at each time interval. Plots for divergence and velocity are shown in Fig. 7 for all three substrates at times 5 minutes after cell division, and 15, 30, and 60 minutes before cell division. These results show a small negative divergence at the division site prior to cell division, demonstrating contraction, and a large positive divergence at the division site at the onset of a divided cell which occurs on all substrates. Following division, regions surrounding the division site contract and cell velocities align in opposite directions. These observations were also independent of substrate stiffness. Rossen and co-workers[31] tracked endothelial cells around a division site and observed similar patterns of contraction and expansion prior to and after cell division. Cell divisions also induced noise and created disorder in the cell velocities in epithelial monolayers[30,32].

We next explored the role of density fluctuations in leader cell formation in our simulations. We define the relative density of follower cells in a region as the ratio of the local density with reference density for each of the substrate stiffness in our study (Table 2). The local density is a function of the mean distance between a cell and its neighbors. A higher mean distance results in a lower value of the local density. Fig. 8 shows density plots at 15, 30, 60 and 120 minutes before the time a given cell assumes the role of a leader cell for all three substrates in our study. These plots show an increase in the density of follower cells that is visible about 2 hours before a cell becomes a leader cell. Regions of higher cell density appear as projections and are located directly behind the leader cell. In supplementary video 3, regions of increased density corresponding to leader cell formation, are observed throughout the migration duration on the 9.4 kPa substrate.

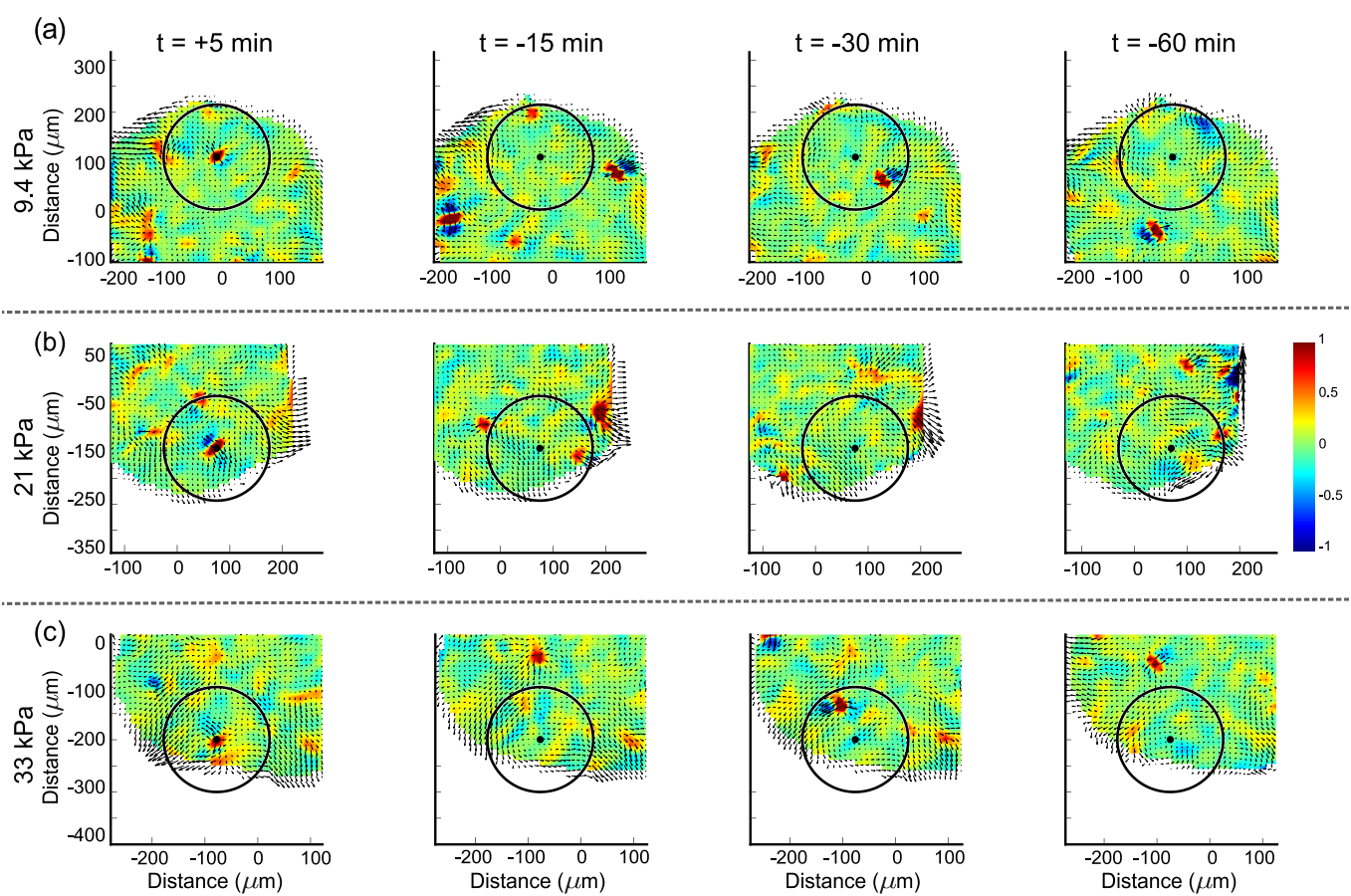

**Figure 7:** Plots of divergence and cell velocities are shown for the (a) 9.4 kPa substrate, (b) 21 kPa substrate, and (c) 33 kPa substrate at 5 mins after cell division and 15 mins, 30 mins and 60 mins before cell division. The divergence indicates whether a particular region in the monolayer is expanding or contracting.

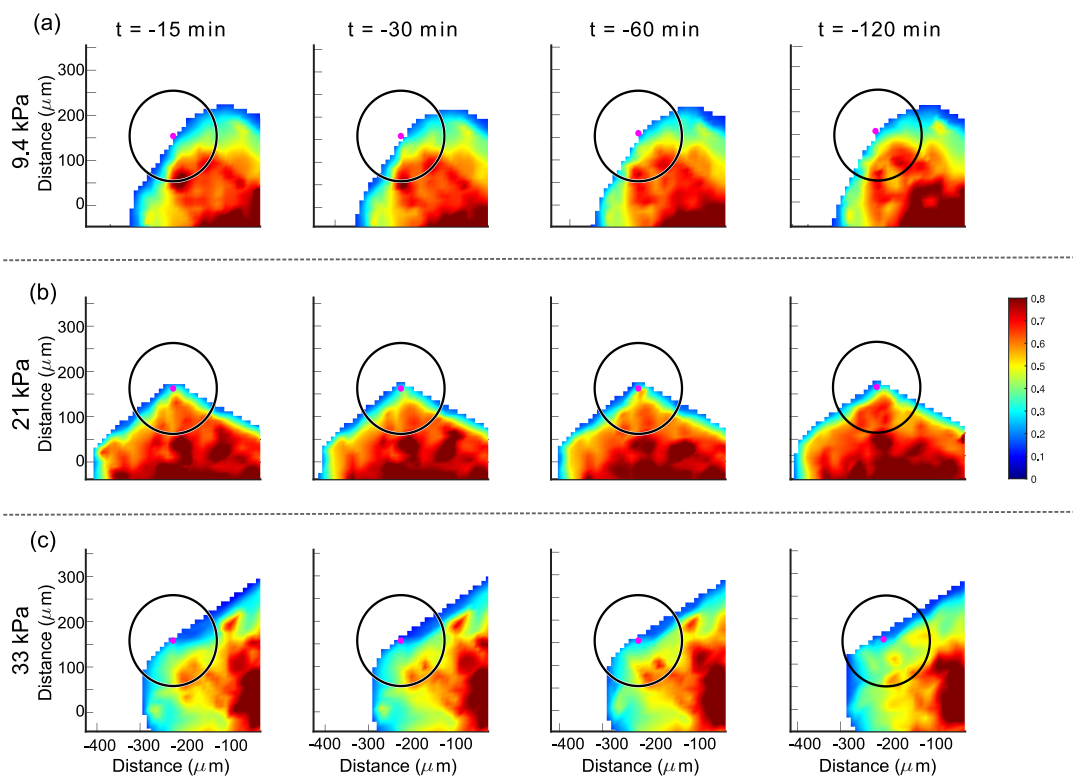

**Figure 8:** Regions of increased density occur behind a cell prior to its emergence as a leader. Plots of relative density for the (a) 9.4 kPa substrate, (b) 21 kPa substrate and (c) 33 kPa substrate are shown at 15 mins, 30 mins, 60 mins and 120 mins prior to a cell (magenta) becoming a leader. The circle defines a 100 µm region around the concerned cell (magenta).

Earlier studies hypothesized that heterogeneities in tractions and velocities in follower cells behind a potential leader induce finger formation at the border of the monolayer[9,17]. Using a continuum model, Alert and co-workers concluded that a velocity gradient from the border to the bulk of the monolayer was sufficient to induce fingering instabilities in a monolayer[9]. Vishwakarma and co-workers also reported increased tractions before the formation of leaders[17]. Results from our study show that fluctuations in cell densities within the monolayer cause heterogeneities in the dynamics of follower cell velocities. These density instabilities may introduce heterogeneities among follow cells that lead to cell polarizations in the border regions. Density fluctuations hence play a direct role in the emergence of leader cells in epithelial monolayers during collective migrations unlike those due to velocity ordering.

## Conclusions

Leader cells are integral in collective cell migrations and help determine the direction of migration. Earlier studies exploring mechanisms of leader cell formation in epithelial monolayers presented two contrasting ideas: first, leader cells are unique and endowed with special attributes[6]. Second, leader cells are created by follower cells[17,27]. We used a particle-based model to explore the role of border and cable forces in the formation of leader cells during collective cell migration. Our results show that border forces, $F_{max}$, were critical in the formation of leader cells and the areal expansion of the monolayer for a given substrate; these results agree with experimental results on substrates of varied stiffness. Simulations show that the border forces drive leader cell formation in the monolayer. Interaction potential terms in the model are also critical determinants of cell-cell interactions which are important in the emergence of a high percentage of leader cells, and help prevent cell detachments. The cable force, $F_{cable}$, and interaction potential terms govern the areal expansion of the monolayer.

Results show that the applied border force induces ordering among the follower cells through Vicsek interactions. This ordering takes place prior to leader cell formation and is a substrate stiffness dependent effect. Cell divisions also play an important role in influencing the velocity profiles at division sites prior to and after cell division. We show that regions of increased local density in follower cells, located within a 100 $\mu$m region behind a leader cell, occur irrespective of the substrate stiffness. These density fluctuations may cause dynamic heterogeneities among follower cells that help form leader cells. Together these results demonstrate that although border forces may not directly create leader cells, they play an important role in collective epithelial migrations. Future studies on the relationships between density fluctuations and monolayer tractions may aid in a better understanding of the mechanisms leading to leader cell formation in epithelial monolayers.

## Author Contributions

SR performed simulations, analyzed results, and helped write the manuscript. SD helped perform simulations. NG designed the study, analyzed results, and wrote the manuscript.

## Conflicts of interest

There are no conflicts to declare.

## Acknowledgements

NG acknowledges financial support from the Department of Biotechnology (BT/PR23724/BRB/10/1606/2017). We also thank Dr. Vijayanand and Prof. Gaurav Tomar for discussions during early part of this work.

## Notes and references